# Harmonics of the AC susceptibility as probes to differentiate the various creep models

*M.G. Adesso[*], C. Senatore, M. Polichetti, S. Pace*

Dipartimento di Fisica, Universita' degli Studi di Salerno-INFM

Via S. Allende, 84081, Baronissi (SALERNO) – ITALY

**Abstract**

We measured the temperature dependence of the 1st and the 3rd harmonics of the AC magnetic susceptibility on some type II superconducting samples at different AC field amplitudes, $h_{AC}$. In order to interpret the measurements, we computed the harmonics of the AC susceptibility as function of the temperature T, by integrating the non-linear diffusion equation for the magnetic field with different creep models, namely the vortex glass-collective creep (single-vortex, small bundle and large bundle) and Kim-Anderson model. We also computed them by using a non-linear phenomenological I-V characteristics, including a power law dependence of the pinning potential on $h_{AC}$. Our experimental results were compared with the numerically computed ones, by the analysis of the Cole-Cole plots. This method results more sensitive than the separate component analysis, giving the possibility to obtain detailed information about the contribution of the flux dynamic regimes in the magnetic response of the analysed samples.

## 1. Introduction

The dynamic magnetic processes in many type II superconductors (both the HTS and the $MgB_2$) are mainly ascribed to thermally activated phenomena, namely the flux creep.

The first creep model was the Kim-Anderson one, in which a linear dependence of the pinning potential U on the current density J is supposed [1]. Later, different creep models have been introduced, such as the vortex glass-collective creep [2,3,4], with strongly non-linear relations between U and J, in order to explain some controversial experimental data. Aim of this work is to show that the existing creep models are not enough complete to explain the experimental data of fundamental and higher harmonics of the AC susceptibility, at least in YBCO and in the $MgB_2$ samples. These discrepancies between the creep theories and the experimental data are more evident if we consider the Cole-Cole plots of the harmonics instead of each separate component. Nevertheless, if we use a phenomenological resistivity model [5], derived by the study of I-V characteristics, we can reproduce some peculiar experimental behaviours.

---

[*] Corresponding author. Tel.: ++39-89-965 256; fax: ++39-89-965 275; e-mail: adesso@sa.infn.it.

## 2. Experimental results

We measured the 1$^{st}$ $(\chi_1', \chi_1'')$ and the 3$^{rd}$ $(\chi_3', \chi_3'')$ harmonics of the AC susceptibility on both melt-grown YBCO and MgB$_2$ slabs, as a function of the temperature, at different amplitudes of the AC magnetic field, $h_{AC}$, at a fixed frequency ν and without a DC field. In Fig.1 and in Fig.2 the Cole-Cole plots of the 1$^{st}$ and 3$^{rd}$ harmonics are shown respectively for an YBCO and an MgB$_2$ sample.

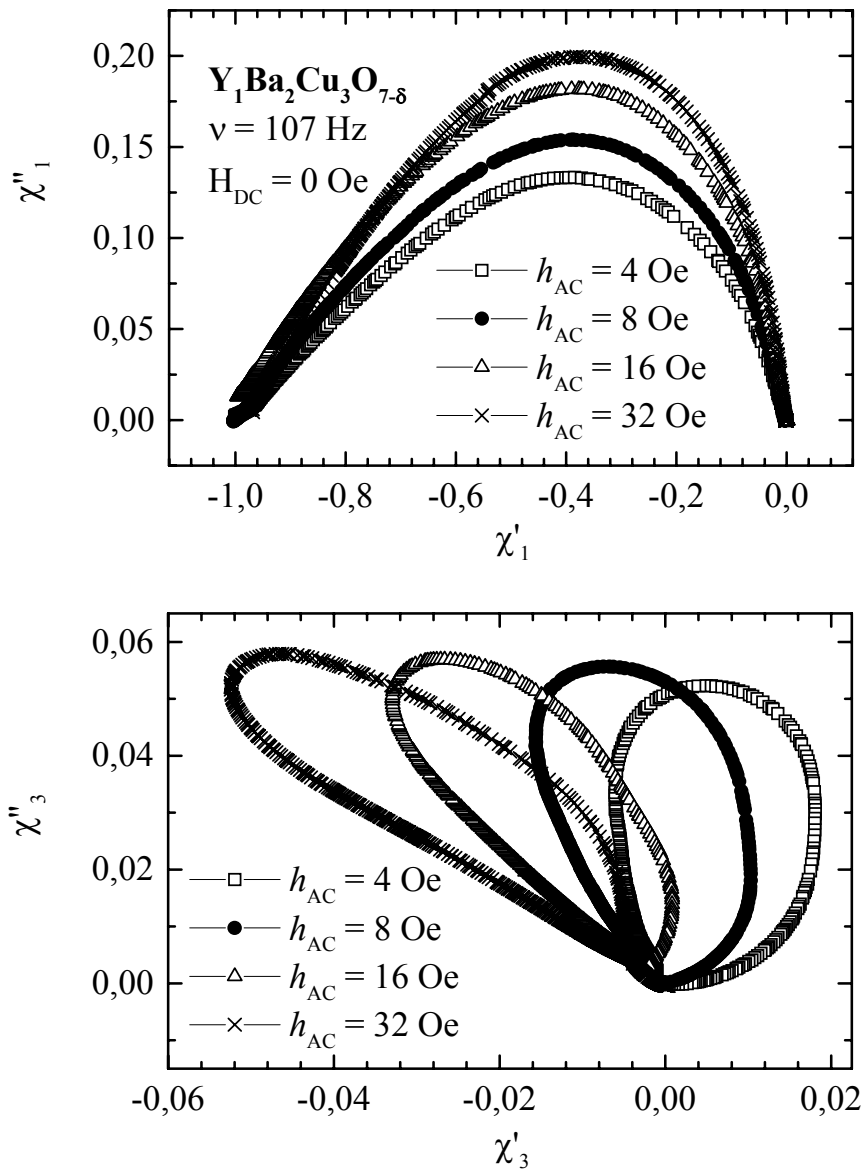

Fig.1: Cole-Cole plots of 1$^{st}$ and 3$^{rd}$ harmonics of the AC susceptibility on a YBCO sample, at different $h_{AC}$.

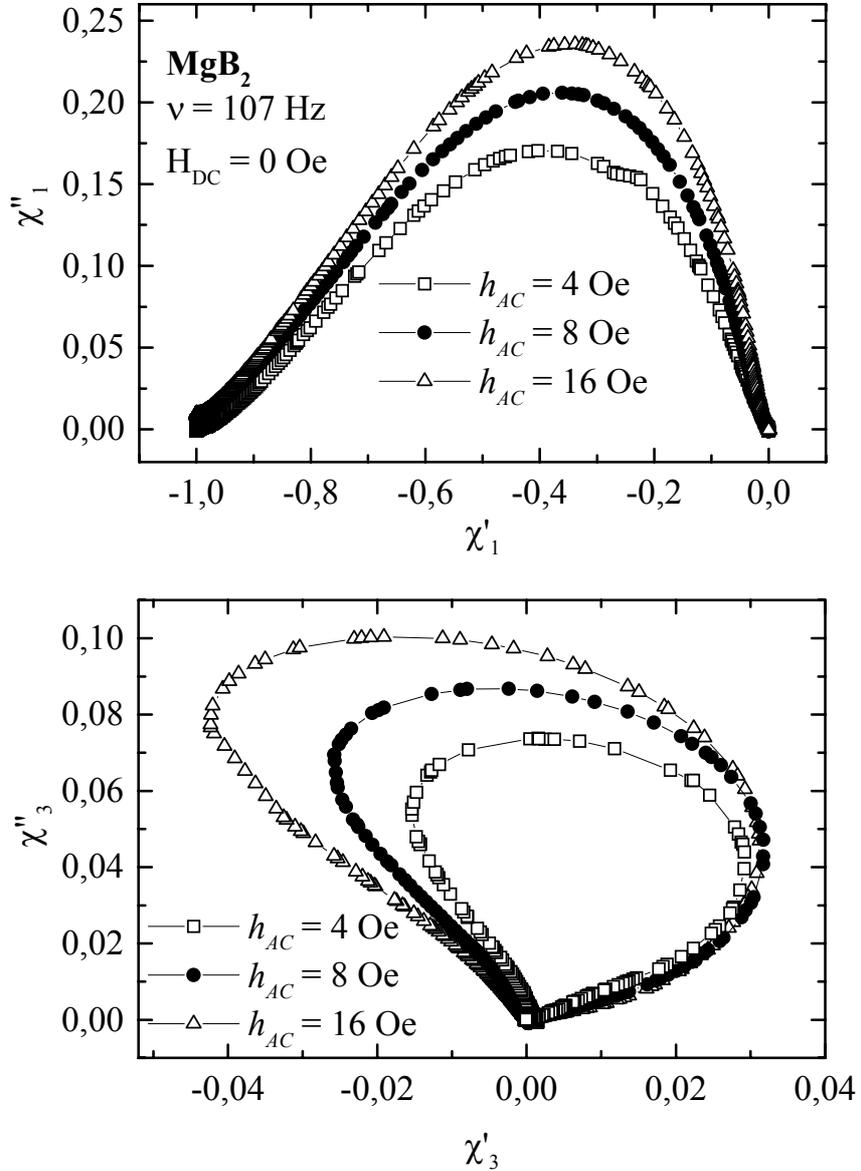

Fig.2: Cole-Cole plots of 1$^{st}$ and 3$^{rd}$ harmonics of the AC susceptibility on a MgB$_2$ sample, at different $h_{AC}$.

We can observe that, in both the samples, the Cole-Cole plots of the 1$^{st}$ harmonics are characterized by a dome-shaped curve with a maximum that grows if $h_{AC}$ is increased. Also the 3$^{rd}$ harmonics have a common feature: in both the materials the Cole-Cole plots tend to occupy the left semi-plane more and more, if $h_{AC}$ is increased. Nevertheless, for the YBCO the area of the $\chi_3$ Cole-Cole loops is constant within 1%, whereas in the MgB$_2$ it increases when $h_{AC}$ grows.

# 3. Numerical results

In order to interpret our data, we compared them with numerical curves obtained by integrating the non-linear diffusion equation of the magnetic field inside the samples. In all the simulations, we chose the temperature dependence of the pinning potential according to the δl-type collective pinning model, as it has been experimentally verified to be valid for both the samples [5,6]. By changing the dependence U(J) of the pinning potential on the current, we analysed different models.

*3.1 Creep models*

The Vortex Glass theory predicts that the pinning energy has a non-linear dependence on the current density, through the $\mu$ parameter. We use, according to several authors [7], the following model:

$$U_p(J) = \frac{U_0}{\mu}\left[\left(\frac{J_C}{J}\right)^\mu - 1\right] \qquad (1)$$

In the framework of the Collective-Creep theory, the $\mu$–parameter has the following values [8]: 1/7 for single vortex; 3/2 for small bundle and 7/9 for large bundle creep. In Figs. 3-4-5 we report the Cole-Cole plots of the 1st and 3rd harmonics, obtained by using the vortex glass-collective creep model in all these regimes. Summarising, in all the analysed creep models, if $h_{AC}$ is increased, the 1st harmonics Cole-Cole plots have a decreasing maximum and the 3rd harmonics tend to occupy the right semi-plane: both these behaviours are not in agreement with the experimental ones.

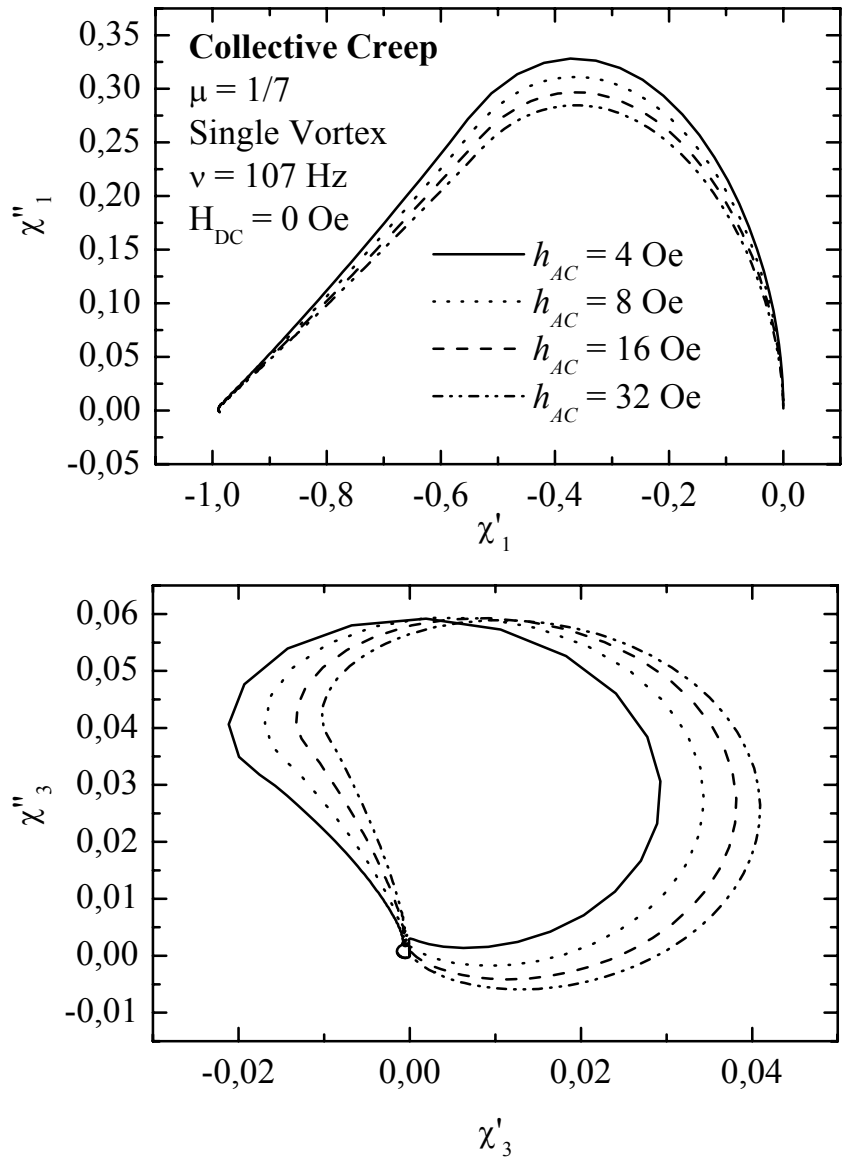

Fig.3: 1st and 3rd harmonics Cole-Cole plots, simulated by using the Vortex glass-Collective Creep model in the single vortex regime, at different $h_{AC}$.

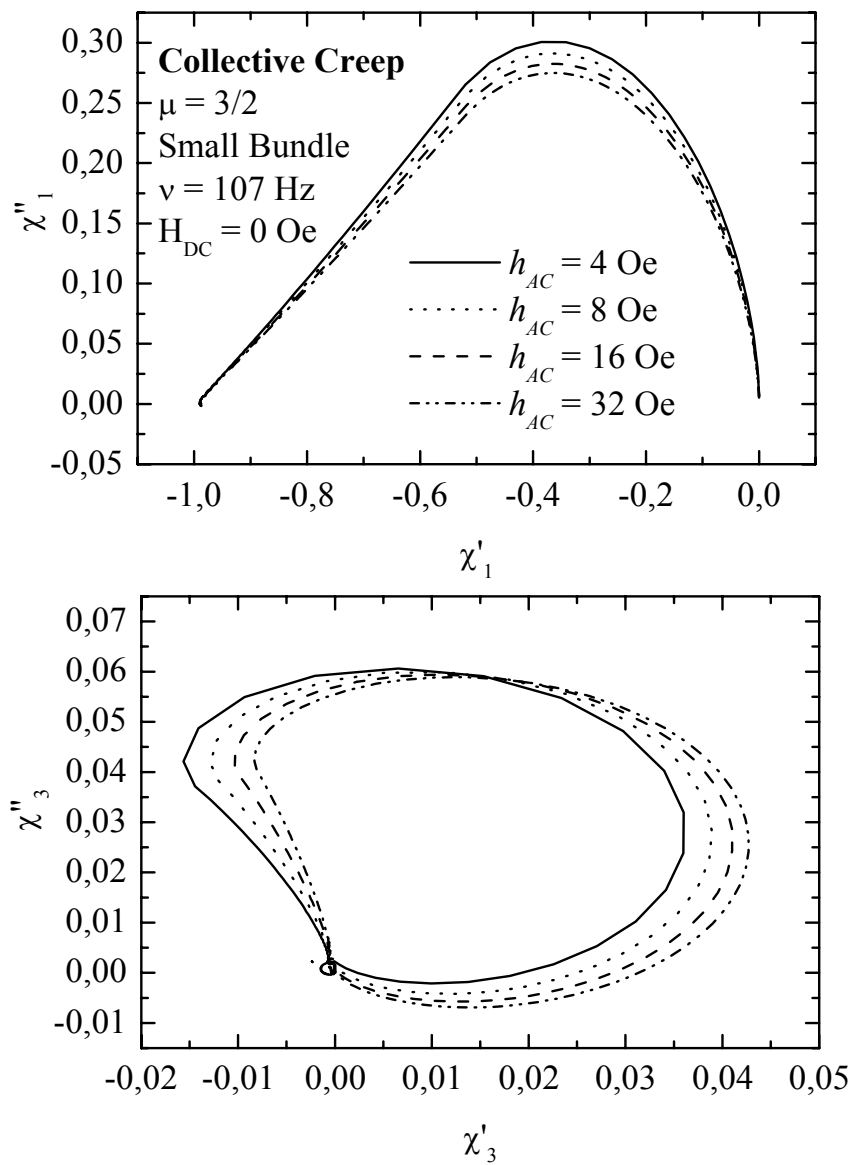

Fig.4: 1st and 3rd harmonics Cole-Cole plots, in the small bundle regime, at different $h_{AC}$.

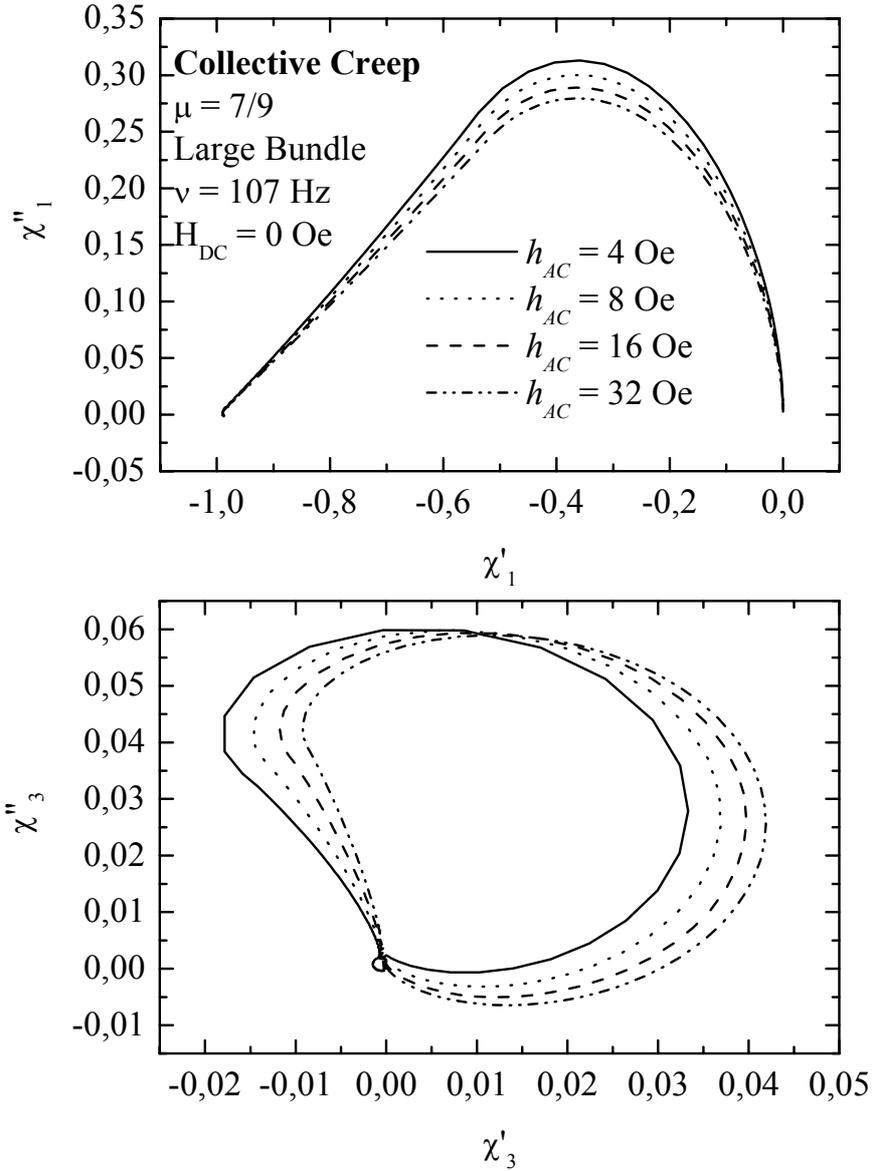

Fig.5: 1st and 3rd harmonics Cole-Cole plots, in the large bundle regime, at different $h_{AC}$.

The (1) with $\mu = -1$ corresponds to the linear Kim-Anderson model [7]. We simulated the 1st and 3rd harmonics Cole-Cole plots in this case, and the results (not reported here) are similar to those in Figs. 3-5.

*3.2 Phenomenological model*

We also used a phenomenological model, in which the pinning potential depends on the current density by a logarithmical law and on $h_{AC}$ by a power law:

$$U_p(J, h_{AC}) \propto \ln(J_c/J) \cdot h_{AC}^{-\beta} \qquad (2)$$

We already showed in a previous work [5] that the simulations obtained with this model have different behaviours depending on the β-value.

Here we report the 1st and the 3rd harmonics Cole-Cole plots, calculated with β=0.645 (Fig.6).

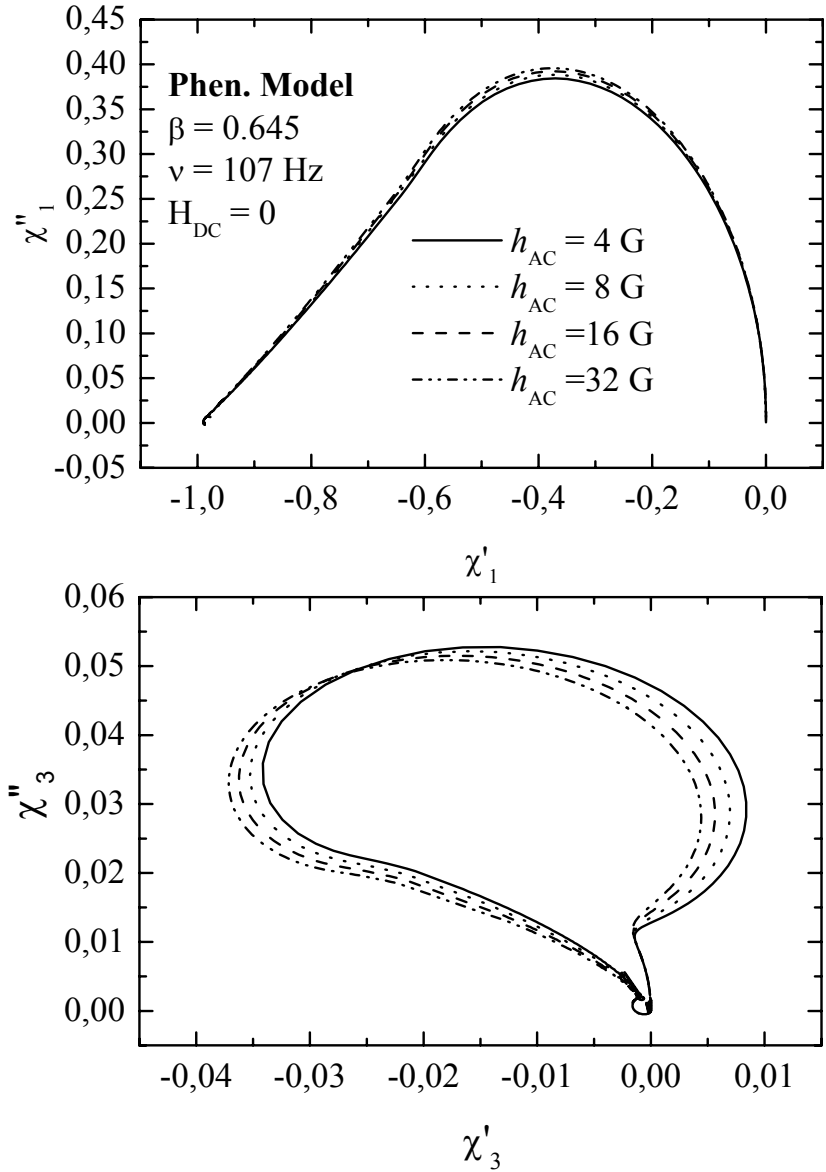

Fig.6: 1st and 3rd harmonics Cole-Cole plots, simulated by using a phenomenological model, with β=0.645, at different $h_{AC}$.

As it is visible in Fig.6, by using this model, we can reproduce the general behaviour of both the samples: an increasing peak in the 1st harmonics Cole-Cole plots, and the 3rd harmonics Cole-Cole loops pointing toward the left semi-plane, if $h_{AC}$ is increased. Nevertheless, there are some discrepancies with the experimental data, in particular with the field dependence of the MgB$_2$ 3rd harmonics (Fig.2). They might be reduced by introducing an $h_{AC}$ dependence in the existing creep models. In this way we could also verify if the agreement with the results of the phenomenological model is due to the particular current dependence used or to the introduction of $h_{AC}$ in the pinning potential.

## 4. Conclusions

We measured the temperature dependence of the 1$^{st}$ and the 3$^{rd}$ harmonics of the AC magnetic susceptibility both on YBCO and MgB$_2$ samples, at different $h_{AC}$. The experimental data were analysed in terms of Cole-Cole plots and compared with those computed by integrating the non-linear diffusion equation for the magnetic field. In the simulations we chose both a power law dependence of the pinning potential on the current density (corresponding to the vortex glass-collective creep) and a logarithmical dependence on the current, including, in this last case, a power law dependence on $h_{AC}$. We observed that for increasing $h_{AC}$ there are some common features in the magnetic response of both the superconductors (namely a growing maximum in the 1$^{st}$ harmonics and a tendency to occupy the left semi-plane in the 3$^{rd}$ harmonics Cole-Cole plots) and we reproduced them only by using the phenomenological model. Nevertheless, some discrepancies between the experimental and the computed curves are still present, mainly in the 3$^{rd}$ harmonics, and the work is in progress, to try to reduce them by introducing an $h_{AC}$ dependence of the pinning potential in the existing creep models.